\documentclass[twocolumn,traditabstract,letter]{aa}
\usepackage{amssymb}
\usepackage{mathptmx}
\usepackage{natbib}
\bibpunct{(}{)}{;}{a}{}{,}
\usepackage[]{graphicx}
\usepackage{lscape}
\usepackage{txfonts}
\usepackage{verbatim}
\usepackage{url}
\defcitealias{canameras20}{C20}
\defcitealias{sonnenfeld18}{S18}
\defcitealias{sonnenfeld20}{S20}
\defcitealias{wong18}{W18}

\begin{document}

\title{HOLISMOKES -- VI. New galaxy-scale strong lens candidates from the HSC-SSP imaging survey
  \thanks{Table 1 is only available in electronic form at the CDS via anonymous ftp to cdsarc.u-strasbg.fr (130.79.128.5)
or via {\tt http://cdsweb.u-strasbg.fr/cgi-bin/qcat?J/A+A/}.}}

\author{R. Ca\~nameras\inst{1}, S. Schuldt\inst{1,2}, Y. Shu\inst{1,3}, S. H. Suyu\inst{1,2,4}, S. Taubenberger\inst{1}, 
T. Meinhardt\inst{5}, L. Leal-Taix\'e\inst{5}, D. C.-Y. Chao\inst{4}, K. T. Inoue\inst{6}, A. T. Jaelani\inst{7}, A. More\inst{8,9}}

\institute{
Max-Planck-Institut f\"ur Astrophysik, Karl-Schwarzschild-Str. 1, 85748 Garching, Germany \\
{\tt e-mail: rcanameras@mpa-garching.mpg.de}
\and
Technische Universit\"at M\"unchen, Physik Department, James-Franck Str. 1, 85741 Garching, Germany
\and
Ruhr University Bochum, Faculty of Physics and Astronomy, Astronomical Institute (AIRUB), German Centre for Cosmological
Lensing, 44780 Bochum, Germany
\and
Institute of Astronomy and Astrophysics, Academia Sinica, 11F of ASMAB, No.1, Section 4, Roosevelt Road, Taipei 10617,
Taiwan
\and
Technical University of Munich, Department of Informatics, Boltzmann-Str. 3, 85748 Garching, Germany
\and
Faculty of Science and Engineering, Kindai University, Higashi-Osaka, 577-8502, Japan
\and
Astronomy Research Group and Bosscha Observatory, FMIPA, Institut Teknologi Bandung, Jl. Ganesha 10, Bandung 40132,
Indonesia
\and
Kavli Institute for the Physics and Mathematics of the Universe (WPI), UTIAS, The University of Tokyo, Kashiwa, Chiba 277-8583, Japan
\and
The Inter-University Centre for Astronomy and Astrophysics (IUCAA), Post Bag 4, Ganeshkhind, Pune 411007, India
}

\titlerunning{HOLISMOKES -- VI. HSC lens search}

\authorrunning{R. Ca\~nameras et al.} \date{Received / Accepted}

\abstract{We have carried out a systematic search for galaxy-scale strong lenses in multiband imaging from the Hyper Suprime-Cam (HSC)
  survey. Our automated pipeline, based on realistic strong-lens simulations, deep neural network classification, and visual inspection,
  is aimed at efficiently selecting systems with wide image separations (Einstein radii $\theta_{\rm E} \sim 1.0$--3.0\arcsec),
  intermediate redshift lenses ($z \sim 0.4$--0.7), and bright arcs for galaxy evolution and cosmology. We classified $gri$ images of
  all 62.5 million galaxies in HSC Wide with $i$-band Kron radius $\geq$0.8\arcsec\ to avoid strict preselections and to prepare for
  the upcoming era of deep, wide-scale imaging surveys with Euclid and Rubin Observatory. We obtained 206 newly-discovered candidates
  classified as definite or probable lenses with either spatially-resolved multiple images or extended, distorted arcs. In addition, we
  found 88 high-quality candidates that were assigned lower confidence in previous HSC searches, and we recovered 173 known systems in
  the literature. These results demonstrate that, aided by limited human input, deep learning pipelines with false positive rates as low
  as $\simeq$0.01\% can be very powerful tools for identifying the rare strong lenses from large catalogs, and can also largely extend
  the samples found by traditional algorithms. We provide a ranked list of candidates for future spectroscopic confirmation.}


\maketitle

\section{Introduction}
\label{sec:intro}

Strong gravitational lensing systems are very powerful tools for probing galaxy evolution and cosmology. They provide constraints to
the level of a few percent on the total mass of the foreground galaxies or galaxy clusters producing the light deflections
\citep[e.g.,][]{bolton08,shu17,caminha19}. This leads to unique diagnostics on the dark matter mass distributions, and to test galaxy
evolution models and the flat Lambda cold dark matter ($\Lambda$CDM) cosmological model. Moreover, strongly lensed time-variable sources
with observed time delays between multiple images provide independent and competitive measurements of the Hubble constant $H_{\rm 0}$
\citep[e.g.,][]{refsdal64,wong20}. However, identifying statistical samples of strong lenses remains a major challenge.

Convolutional neural networks \citep[CNNs;][]{lecun98} have proven extremely efficient for pattern recognition tasks and have given a
strong impetus to image analysis and processing. Recent studies largely demonstrate the ability of supervised CNNs to identify the
rare gravitational lenses among large datasets \citep[e.g.,][]{jacobs17,jacobs19b,petrillo19a,huang20}, extending previous automated
algorithms \citep[e.g.,][]{gavazzi14,joseph14} generally with better classification performance \citep{metcalf19}. In \citet[][hereafter
  \citetalias{canameras20}]{canameras20}, we show that realistic simulations and careful selection of negative examples are crucial for
successfully conducting a systematic search over 30\,000\,$\deg^2$ with PanSTARRS multiband imaging.

We develop here new supervised neural networks for automated selection of galaxy-scale strong lenses in large-scale multiband surveys,
using the Hyper-Suprime Cam Subaru Strategic Program \citep[HSC-SSP;][]{aihara18a} for testing on images approaching the expected depth
and quality of the Rubin Observatory Legacy Survey of Space and Time (LSST) final stacks \citep[see][]{ivezic19}. Previous non-machine
learning identification of galaxy-, group-, and cluster-scale lenses from the Survey of Gravitationally lensed Objects in HSC Imaging
\citep[SuGOHI;][hereafter \citetalias{sonnenfeld18}, \citetalias{sonnenfeld20}, \citetalias{wong18} for SuGOHI-g]{sonnenfeld18,
  sonnenfeld19,sonnenfeld20,wong18,chan19,jaelani20,jaelani21} offers an independent observational set to test our classification
completeness. In this letter we validate our deep learning pipeline and present new high-confidence galaxy-scale lens candidates with
wide image separations from the Wide layer of the HSC survey, which is only $\simeq$1~mag shallower than LSST ten-year stacks, and has
sufficient sky coverage for lens searches. In Section~\ref{sec:method} we describe the neural network and the datasets used for
training, validating, and testing the network, and for searching for new lenses. The classification procedure is presented in
Section~\ref{sec:visu}, and the results are discussed in Section~\ref{sec:results}.

\section{Methodology}
\label{sec:method}

We conducted our search on $gri$ cutouts from HSC Wide public data release 2 (PDR2) covering nearly 800\,$\deg^2$ \citep{aihara19} in
all three bands, out of the final 1400\,$\deg^2$. In PDR2, about 300\,$\deg^2$ reach the nominal 5$\sigma$ point-source sensitivities
of 26.8, 26.4, and 26.2~mag respectively in $g$, $r$, and $i$. We focused on systems with luminous red galaxies (LRGs) acting as lenses
and with image separations $\gtrsim$0.75\arcsec, larger than the median seeing FWHMs in $g$, $r$, and $i$ bands. Such systems are ideal
for constraining the lens mass-density profiles and for finding strongly lensed supernovae for cosmography and stellar physics, which are
major goals of our ongoing Highly Optimized Lensing Investigations of Supernovae, Microlensing Objects, and Kinematics of Ellipticals
and Spirals \citep[HOLISMOKES;][]{suyu20}. To go beyond previous studies that relied on strict catalog preselections, we demonstrate
that a dedicated neural network trained on a carefully constructed dataset can automatically and efficiently identify such lenses
over an extended galaxy sample. We focused on the 62.5 million galaxies observed in $gri$ bands in PDR2, without flagged artifacts,
and with $i$-band Kron radius $\geq$0.8\arcsec. We used 12\arcsec\,$\times$\,12\arcsec\ cutouts, sufficient for galaxy-scale lenses,
downloaded from the data archive server \citep{bosch18}. The design of the dataset and choice of network architecture resulted from
thorough tests of classification completeness and purity, using the test set described in Sect.~\ref{ssec:test}. The performance of
our different networks will be compared in a future paper (Ca\~nameras et al., in prep.).

\subsection{Constructing the ground truth dataset}
\label{ssec:dataset}

\begin{figure}
\centering
\includegraphics[width=.49\textwidth]{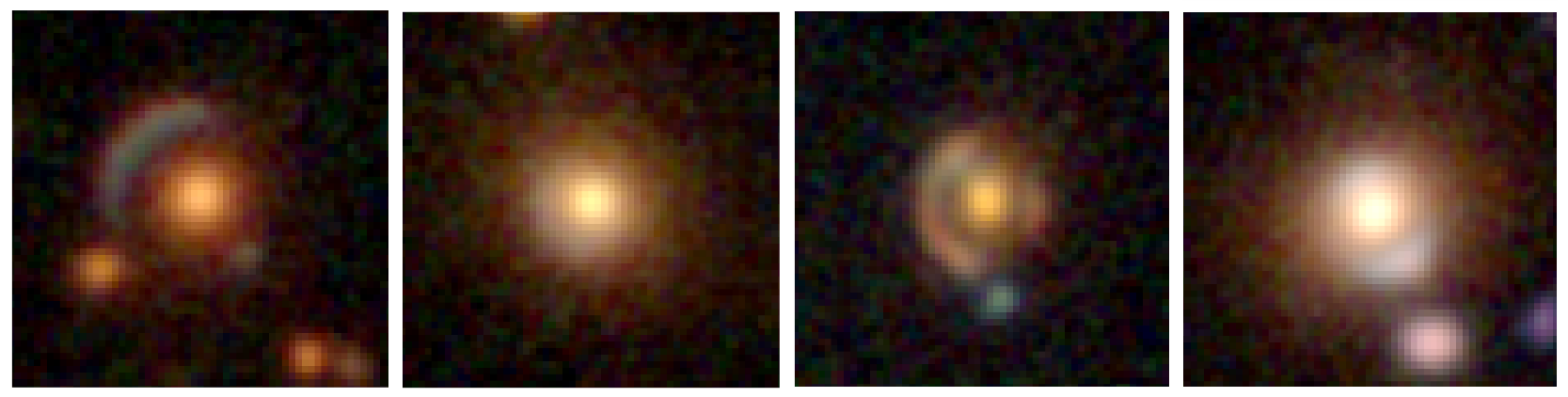}
\includegraphics[width=.49\textwidth]{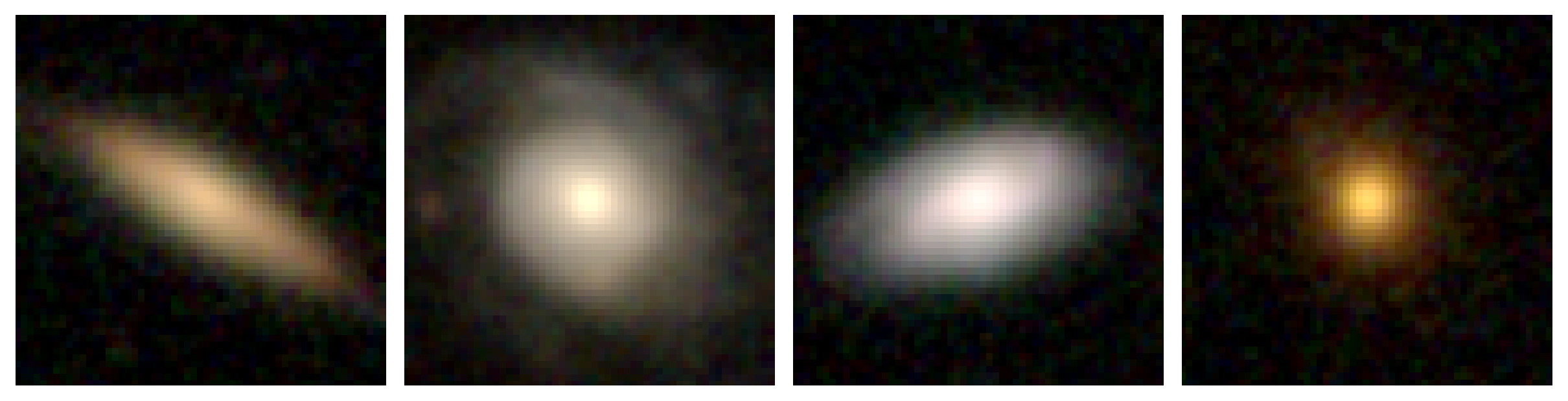}
\caption{Examples of positives (mock lenses, {\it top row}) and negatives (spirals, LRGs, and random nonlenses, {\it bottom row}) in the
  ground truth dataset for training and validation. Each cutout is 12\arcsec\,$\times$\,12\arcsec.}
\label{fig:dataset}
\end{figure}

Supervised machine learning classification depends strongly on the construction of the ground truth data used by the network to learn
the morphological features relevant to each class. We trained and validated our binary classification network with a balanced set of
40\,000 positive and 40\,000 negative examples (Fig.~\ref{fig:dataset}) obtained from random sky positions to limit biases from
small-scale seeing and depth variations. The GAMA09H field was excluded and reserved for a future comparison study of various
lens search pipelines (More et al., in prep.).

As positive examples, we produced realistic galaxy-scale lens simulations by painting lensed arcs on HSC $gri$ images of LRGs. This
approach accounts for the quality of HSC imaging and for the presence of artifacts and neighboring galaxies. We followed the procedure
described in \citet{schuldt21a} and \citetalias{canameras20}, by modeling the lens mass distributions with Singular Isothermal Ellipsoids
(SIE) using LRG redshifts and velocity dispersions from SDSS, and inferring axis ratios and position angles from the light profiles.
Unlike in \citetalias{canameras20} we included external shear, and we chose lens--source pairs to produce a uniform Einstein radius
distribution in the range 0.75\arcsec$-$2.5\arcsec, increasing the number of wide separations and of fainter ($z>0.7$) lens galaxies to
help recover these configurations. As background sources, we used high S/N galaxies with spectroscopic redshifts from the {\it Hubble}
Ultra Deep Field \citep{inami17}, applying color corrections that match HST filter passbands to HSC, and a common flux boost to the
three bands. Sources were lensed with {\tt GLEE} \citep{suyu10,suyu12}, convolved with the PSF model at the location of the lens from
the HSC archive, and coadded with the lens HSC cutout. Mocks that have lensed images with $\mu \geq 5$, $S/N > 5$, and that are brighter
than the lens at the position of peak lensed image emissions are accepted by the pipeline. We used similar fractions of quadruply- and
doubly-imaged systems.

\begin{figure}
\centering
\includegraphics[width=.50\textwidth]{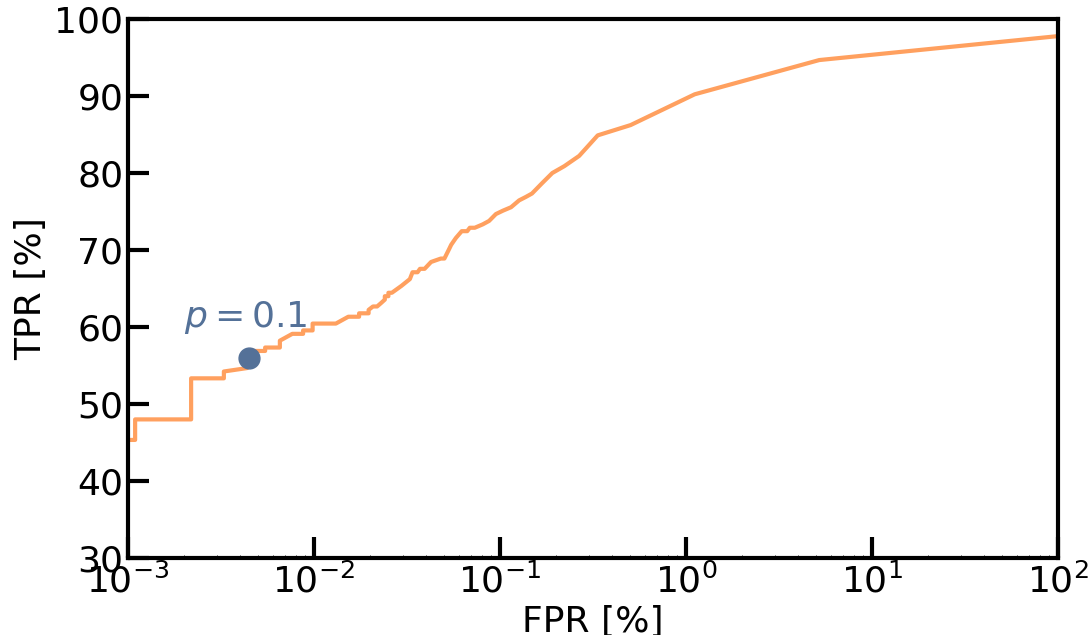}
\caption{Receiver Operating Characteristic curve for our ResNet using an independent test set from HSC Wide PDR2 survey data. Different
  threshold scores of the ResNet trace out the orange curve, and the threshold score of $p=0.1$ is indicated by the blue dot. The true
  positive rate (TPR) corresponds to the number of SuGOHI galaxy-scale test lenses correctly classified over the 202 test lenses. The
  false positive rate (FPR) is measured using random nonlenses in the COSMOS field with Kron radius larger than 0.8\arcsec\ and is
  defined as the number of nonlenses identified as lenses over the total number of nonlenses.}
\label{fig:roc}
\end{figure}

\begin{table*}
  \caption{High-confidence galaxy-scale lens candidates with ResNet scores $>0.1$, and average grades $\geq 1.5$ from visual inspection.}
\centering
\begin{tabular}{lccccccccccc}
\hline
\hline \\[-0.3em]
Name & RA & Dec & $p$ & $G$ & $\sigma_{\rm G}$ & $\# G$=3 & $g_{\rm Kron}$ & $r_{\rm Kron}$ & $i_{\rm Kron}$ & $z$ & Notes \\[+0.5em]
\hline \\[-0.3em]
HSCJ0102$+$0158  &  15.6597  &  1.98247  &  0.12  &  3.0  &  0.0  &  5  &  27.61  &  21.62  &  20.58  &  0.8669$^{(*)}$  &  (i)  \\
HSCJ0157$-$0330  &  29.3812  &  $-$3.51603  &  0.16  &  3.0  &  0.0  &  5  &  21.9  &  21.02  &  20.14  &  0.6212$^{(*)}$  &  (a)  \\
HSCJ0200$-$0344  &  30.1981  &  $-$3.73763  &  0.6  &  3.0  &  0.0  &  5  &  22.23  &  21.17  &  20.04  &  0.72  &    \\
HSCJ0232$-$0323  &  38.2078  &  $-$3.39058  &  0.19  &  3.0  &  0.0  &  5  &  20.89  &  19.37  &  18.62  &  0.42  &  (h) (e) (i)  \\
HSCJ0236$-$0332  &  39.1554  &  $-$3.53893  &  1.0  &  3.0  &  0.0  &  5  &  20.48  &  19.09  &  18.54  &  0.2695$^{(*)}$  &  (a) (e)  \\
HSCJ0238$-$0545  &  39.5741  &  $-$5.76545  &  0.72  &  3.0  &  0.0  &  5  &  20.75  &  19.96  &  19.18  &  0.5993$^{(*)}$  &  (a) (c)  \\
HSCJ0850$+$0039  &  132.6942  &  0.65146  &  0.61  &  3.0  &  0.0  &  5  &  24.66  &  22.08  &  21.25  &  1.00  &  (d)  \\
HSCJ0904$-$0059  &  136.0331  &  $-$0.99807  &  0.69  &  3.0  &  0.0  &  5  &  22.42  &  21.23  &  20.26  &  0.56  &  (k)  \\
\multicolumn{12}{l}{{\it Table continues as Supplementary material that is available in the online version of the paper.}} \\
\hline
\end{tabular}
\tablefoot{Some systems in this table are also found in separate lens searches and cross-references will be added to the corresponding
  publications (Shu et al., in prep.; Jaelani et al., in prep.). Columns are: source name; right ascension; declination; output score
  from the ResNet; average visual grades from five authors; dispersion in the grades; number of classifiers assigning the highest grade of
  {\tt 3}; $g$-, $r$-, and $i$-band Kron magnitudes from PDR2; CNN photometric redshift estimates from \citet{schuldt21b} or spectroscopic
  redshifts marked as $^{(*)}$ where available; references for systems previously published, either as spectroscopically-confirmed lenses
  or as grade A or B candidates. References are the following: (a) \citet{sonnenfeld18}, (b) \citet{wong18}, (c) \citet{chan19}, (d)
  \citet{sonnenfeld20}, (e) \citet{jaelani20}, (f) \citet{canameras20}, (g) \citet{huang19}, (h) \citet{stark13}, (i) \citet{jacobs19b},
  (j) \citet{li20}, (k) \citet{more12}, (l) \citet{petrillo19a}, (m) \citet{brownstein12}, (n) \citet{gavazzi14}, (o) \citet{diehl17},
  (p) \citet{more16}, (q) \citet{shu16}, (r) \citet{jacobs17}, (s) \citet{ratnatunga95}, (t) \citet{tanaka16}, and (u) \citet{more17}.
  Candidates marked with a $\dagger$ have a lower grade C in SuGOHI.}
\label{tab:cand}
\end{table*}

As negative examples, we selected a sample of spirals, isolated LRGs, and random galaxies with $r_{\rm Kron}<23$\,mag in similar
proportions, and a few compact galaxy groups. We obtained spirals with Kron radius $<$2\arcsec\ from the catalog of \citet[][also
  from HSC Wide]{tadaki20} in order to boost the fraction of examples mimicking lensed arcs. Isolated LRGs helped the network to
learn that lensed arcs are the relevant features, and groups were selected from \citet{wen12}. Other types frequently misclassified
as lenses (e.g., rings, mergers) were more difficult to include due to limited morphological classifications available in the HSC
footprint \citep[e.g.,][]{willett13}.

\subsection{Training the neural network}
\label{ssec:train}

Building upon the success of CNNs, deeper architectures have been developed to optimize performance such as image classification
accuracies. In particular, the residual learning concept \citep[ResNet;][]{he15} enables one to increase the network depth
and performance without requiring drastic computing resources. Such ResNets have obtained excellent results on the
ImageNet Large Scale Visual Recognition Challenge 2015 \citep[][]{he15}. They resemble deep CNNs with multiple building blocks
\citep[preactivated bottleneck residual units in][]{he16}, and shortcut connections between these blocks that make the
convolutional layers learn residual functions with respect to the previous layer, and help avoid vanishing gradients during
optimization. In the recent past, \citet{lanusse18} have developed ResNet architectures for lens finding on LSST-like simulations,
and obtained better results than classical CNNs on the strong lens finding challenge \citep{metcalf19}. Subsequent studies
confirm that such ResNets can efficiently select lenses on real survey data \citep[e.g.,][]{li20,huang20}.

We used a ResNet adapted from the ResNet-18 architecture \citep{he15} which provides a good trade-off between performance and
total training time for binary classification in ground-based imaging. The network has a total of 18 layers with eight blocks
comprising two convolutional layers with batch normalization and nonlinear ReLU activations. We added a fully connected layer of
16 neurons before the last single-neuron layer with sigmoid activation that outputs a score $p$.

As data augmentation to prevent overfitting and improve generalization, the image centroids were randomly shifted between $-$5 and +5
pixels, negative pixels were clipped to zero, and square root stretch was applied to boost low-luminosity features. Other techniques
such as image normalization did not improve the performance and were thus not used. The dataset was split into 80\% for training and
20\% for validation and, after randomly initializing weights, the ResNet was trained over 100 epochs using mini-batch stochastic
gradient descent with 128 images per batch, a learning rate of 0.0006, a weight decay of 0.001, and a momentum fixed to 0.9. We used
early stopping by saving the final network at epoch 21 that corresponds to the minimal binary cross-entropy loss in the validation
set without overfitting.

\subsection{Testing the performance}
\label{ssec:test}

We used HSC Wide PDR2 images to design a test set closely representative of the overall search sample.
First, the completeness was measured on SuGOHI galaxy-scale lenses that are spectroscopically confirmed or have
grades A or B \citepalias{sonnenfeld18,wong18,sonnenfeld20}. We visually rejected a few lenses with large image separations
$\gtrsim$4\arcsec\ suggesting major perturbation from the lens environment as we do not intend to recover such configurations. Out
of the 220 SuGOHI systems remaining, 202 match our Kron radius $\geq$0.8\arcsec\ threshold and were kept as test lenses. Second, the
expected rate of false positives was automatically measured with a set of nonlens galaxies representative of our overall search
sample, including observational artifacts and various types of interlopers. We collected nonlenses in the COSMOS field
\citep{scoville07}, excluding flagged HSC cutouts and sources with Kron radius lower than 0.8\arcsec, as described above. We
excluded all 130 strong lenses and lens candidates previously listed in the MasterLens database\footnote{http://admin.masterlens.org},
or in \citet{faure08}, \citet{pourrahmani18}, and SuGOHI, assuming that the unparalleled coverage of COSMOS guarantees a nearly
complete lens selection. We then classified $gri$ images of the 91\,000 remaining nonlenses.

As shown in the Receiver Operating Characteristic (ROC) curve (Fig.~\ref{fig:roc}), our ResNet reaches extremely low false positive
rates (FPRs) at least a factor of 10 lower than classical CNNs \citepalias{canameras20}. By drastically limiting the number of contaminants,
this network saves significant human inspection time which makes it very promising for rapid lens finding in any deep, wide-scale imaging
survey. We adopted a ResNet score threshold $p>0.1$ for lens selection to maintain ${\rm FPR \lesssim 0.01}$\% with completeness $>$50\%
in SuGOHI (Fig.~\ref{fig:roc}). A comprehensive discussion on classification accuracies as a function of galaxy properties will be
presented in a future paper, together with our other networks.

Using 6000 COSMOS nonlenses with $r<22$ and the 202 SuGOHI test lenses, we tested the stability of ResNet scores through
few-pixel translations, $k \times \pi/2$ rotations, and flipping of the $gri$ images. Applying 100 random transformations and computing
the output distribution of $p$ showed that predictions with mean $\mu_{\rm p}<0.1$ and $>0.9$ are systematically stable, with a scatter
$\sigma_{\rm p}<0.05$. Galaxies with $p<0.1$ discarded before visual inspection therefore have robust ResNet predictions. Scores with
$\mu_{\rm p}=0.2$--0.8 have higher scatter $\sigma_{\rm p} \simeq 0.05$--0.35.

\section{The classification procedure}
\label{sec:visu}

\begin{figure}
\centering
\includegraphics[width=.48\textwidth]{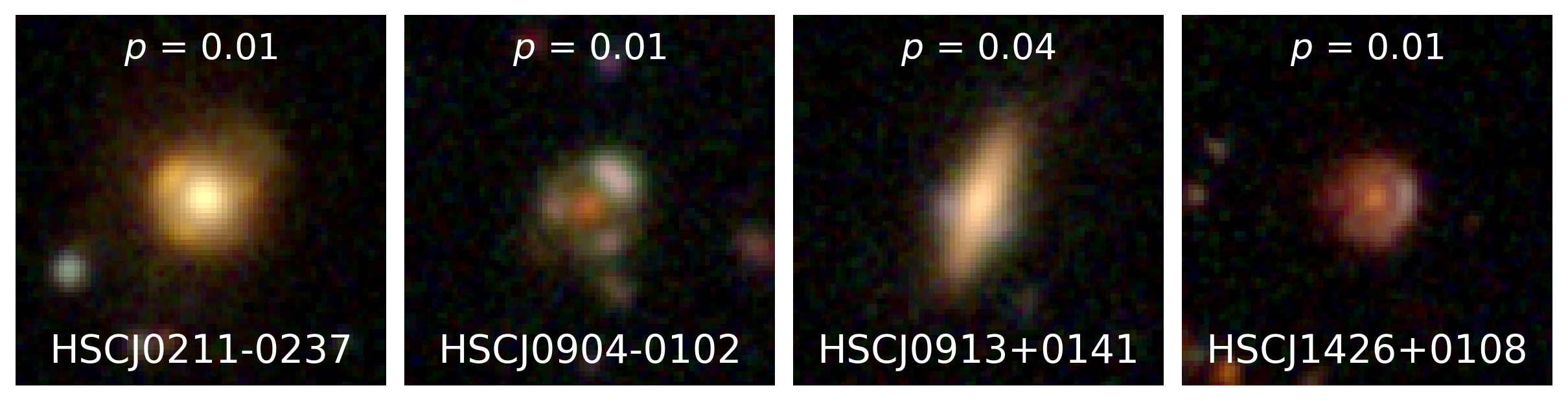}
\includegraphics[width=.48\textwidth]{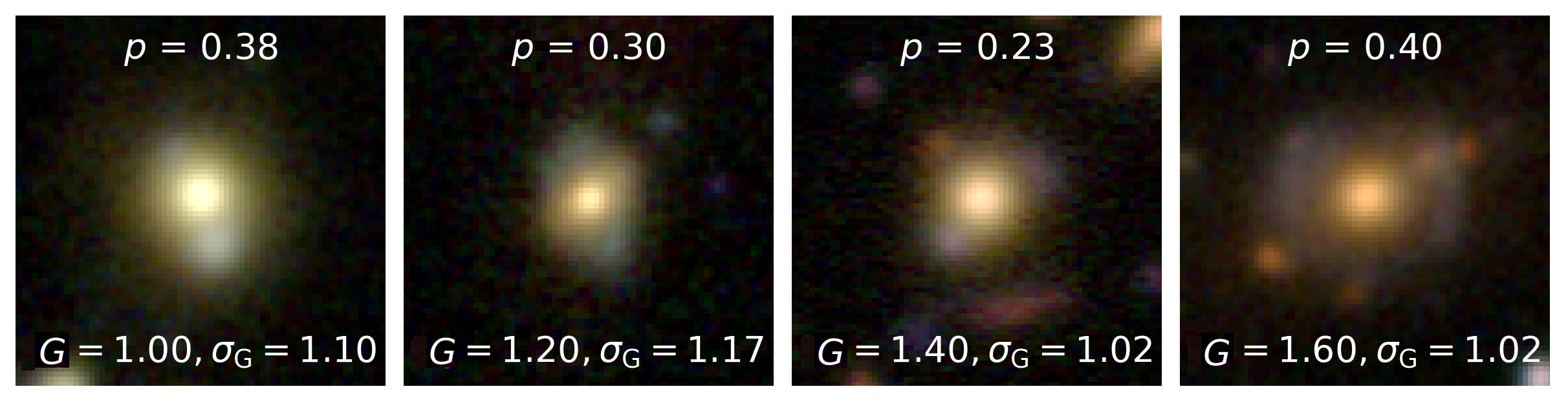}
\includegraphics[width=.48\textwidth]{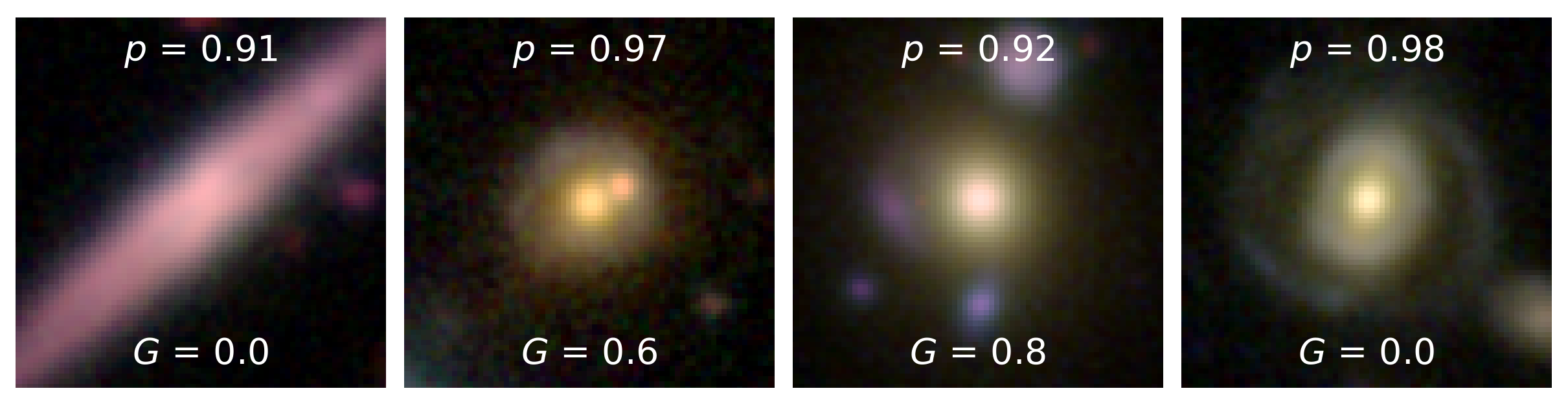}
\caption{Illustration of the network and visual classification stages. {\it Top row:} HSC three-color $gri$ images of some SuGOHI
  galaxy-scale test lenses missed by the neural network, with scores $p<0.1$. {\it Second row:} Subset of ResNet candidates with
  elevated dispersion $\sigma_{\rm G}$ among our visual inspection grades. {\it Third row:} Examples of interlopers with $p>0.1$ and
  low average grades $G$, showing dust lanes or arc-like features around LRGs.}
\label{fig:fp}
\end{figure}

The trained ResNet was applied to the $gri$ cutouts of all 62.5 million galaxies with Kron radius larger than 0.8\arcsec\ in order
to estimate their score $p$. A few hundred cutouts with residual sky background due to imperfect subtraction or to nearby saturated
stars were assigned high scores, and we automatically excluded these cutouts with SExtractor \citep{bertin96}. This resulted in
9651 neural network candidates with $p>0.1$, 0.015\% of the input sample, including 114/202 (56\%) galaxy-scale test lenses from
SuGOHI. We qualitatively observe that the misclassified test lenses (see Fig.~\ref{fig:fp}) tend to have either compact and fainter
lens galaxies, lensed sources with redder colors, stronger blending with lens light, or lower source-to-lens flux ratios. Each of
these configurations is less represented in our simulations. Our ResNet also recovers 102 group- and cluster-scale lens candidates,
although it is not optimized for these systems.

\begin{figure*}
\centering
\includegraphics[width=.98\textwidth]{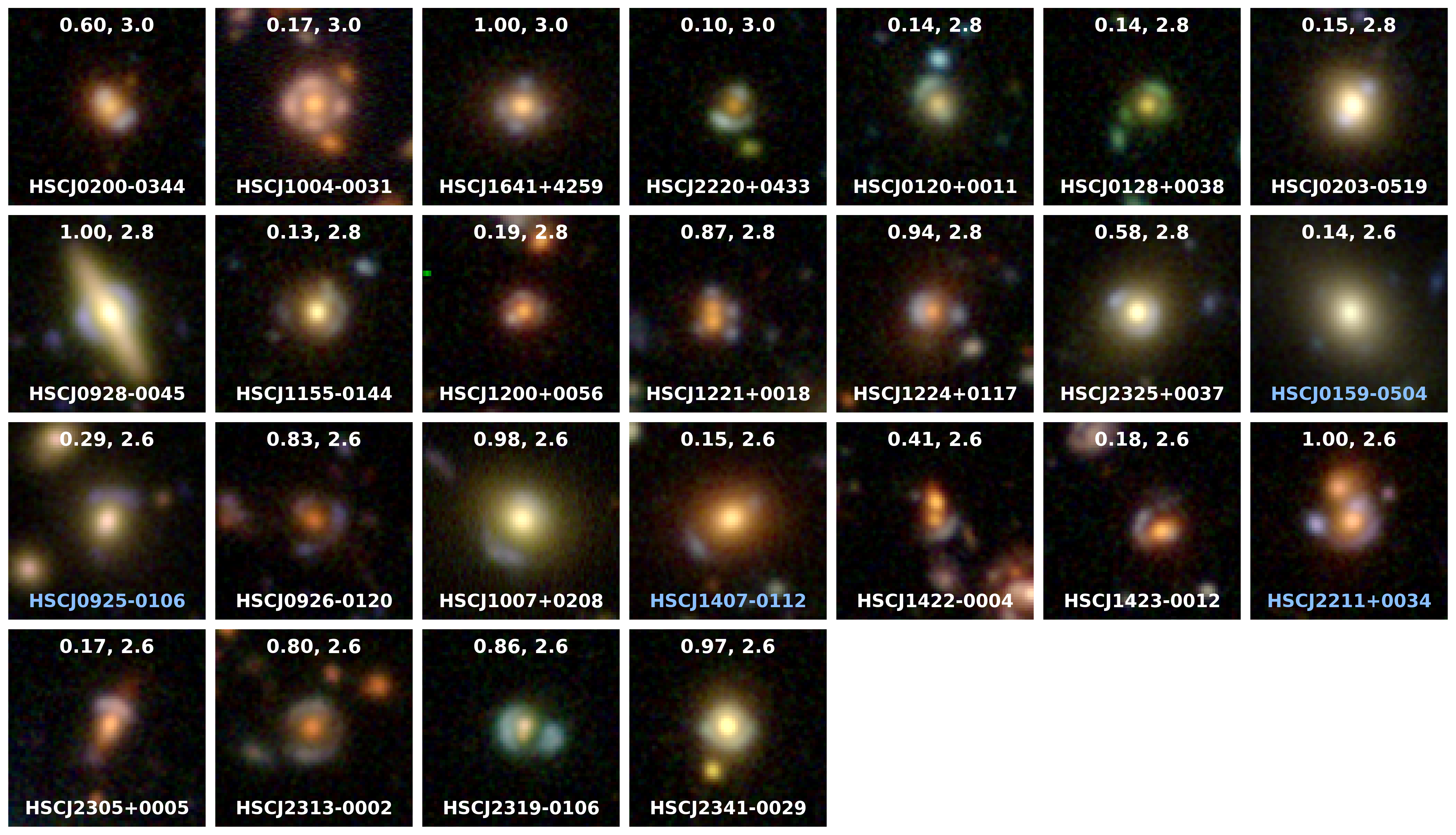}
\caption{Postage stamps (12\arcsec\,$\times$\,12\arcsec) of grade A lens candidates we have discovered in the HSC Wide survey, using
  $gri$ multiband imaging. At the top of each panel we list the ResNet scores $p$, and the average grades $G$ from visual inspection
  of scores $p>0.1$. Grade A corresponds to $G \geq 2.5$. Candidates with white labels are newly discovered as they are not part of
  our compilation of previously confirmed strong lenses and grade A or B lens candidates in the literature. Those marked in light blue
  are listed as grade C in SuGOHI and obtained higher confidence in our classification.}
\label{fig:cnncand}
\end{figure*}
 
The sample with $p>0.1$ contains a large number of false positives, and we conducted a visual inspection stage to collect a final list
of high-confidence lens candidates. Five authors (R.\,C., S.\,S., Y.\,S., S.\,H.\,S., and S.\,T.) inspected three-color images displayed
with different scaling and contrasts, and assigned grades following explicit criteria described in \citetalias{sonnenfeld18} and
\citetalias{canameras20}. In short, grade {\tt 3} corresponds to unambiguous lenses with resolved multiple images, grade {\tt 2}
corresponds to probable lenses with extended and distorted arcs but no obvious counter-image, grade {\tt 1} corresponds to possible
lenses such as LRGs with a single, weakly distorted companion, and grade {\tt 0} includes obvious interlopers such as spirals, mergers,
and galaxy groups. Blind tests using 70 galaxies with $p>0.1$ and 30 SuGOHI lenses led to comparable average grades per classifier, 30\%
of cases with zero dispersion among our grades, and systematic recovery of known lenses, which illustrates the benefit of averaging
individual grades and validates our approach.

While the network scores are not calibrated as probabilities, the fraction of contaminants clearly increases for lower scores. For
this reason, the top 2092 candidates with $p>0.2$ were directly graded by the five authors, while author R.\,C. excluded obvious
interlopers from the 7559 candidates with $0.1<p<0.2$ and forwarded the 739 objects with grades $\geq$1 for inspection by the other
authors. After this first iteration, we reinspected the 332 candidates with dispersion $\geq$0.75 among our five grades. The final
grades have averages of 0.36--0.62 and dispersions of 0.80--0.85 per classifier and were not normalized. Cases with dispersed visual
grades often show ambiguous blue arcs that could either be lensed arcs from background galaxies (without clear counter-images), spiral
arms, or tidal features. The number of grade {\tt 3} lens candidates from each classifier spans a broad range between 69 and 204.
Moreover, our inspection recovers 101/114 SuGOHI test lenses with $p>0.1$.

Most contaminants turn out to be underrepresented in our training set and include edge-on spirals, spirals with diffuse or unresolved
arms discarded from the ``S/Z'' classification of \citet{tadaki20}, lenticular galaxies, and LRGs with dust lanes or with faint and
unlensed companions. In the future, morphological classifications with unsupervised machine learning \citep[e.g.,][]{martin20} or
crowdsourcing will offer interesting avenues for collecting large samples of these galaxy types in the HSC Wide footprint, aiding the
selection of negative examples for supervised lens searches. Image artifacts are already well represented in the training set and were
better excluded.

\section{Results and discussion}
\label{sec:results}

We used the average visual grades $G$ among the five examiners to rank our final sample. We compiled a total of 88 grade A ($G \geq 2.5$)
and 379 grade B ($1.5 \leq G < 2.5$) that have convincing lensing features, corresponding to $\simeq$5\% of network recommendations and
to $\simeq$0.6 candidate per deg$^2$ \citep[close to expectations from simulations by][]{collett15}. Our findings are summarized in
Table~\ref{tab:cand}. The purity, defined as the fraction of grades $G \geq 1.5$ among ResNet recommendations, decreases rapidly when
lowering the threshold $p$. We estimate that 38\% of the highest scores $0.9<p<1.0$ have $G \geq 1.5$, decreasing to 20\% for $0.6<p<0.7$,
7\% for $0.2<p<0.3$, and 3\% for the lowest interval $0.1<p<0.2$.

To find duplicates, this list was cross-matched with our extended compilation of strong gravitational lenses previously published
as confirmed systems or as candidates with confidence levels equivalent to our grades A and B (see \citetalias{canameras20}). Given
the dataset overlap, we extended our cross-match to the full SuGOHI database including grades A, B, and C. We also checked the SIMBAD
Database\footnote{http://simbad.u-strasbg.fr/simbad/sim-fcoo} and the Hubble Source Catalog\footnote{https://catalogs.mast.stsci.edu/hsc/}.
A total of 21/88 grade A and 185/379 grade B lens candidates are newly discovered, and our inspection increases confidence for 4/88 grade
A and 84/379 grade B that were assigned grade C in SuGOHI. A subset of these 294 new high-quality candidates best suited for spectroscopic
follow-up is shown in Fig.~\ref{fig:cnncand}. References of lenses in the literature we recovered are listed in Table~\ref{tab:cand}. We
analyzed SDSS DR16 spectra available for a subset of candidates, and systematically found signatures of LRGs at intermediate redshift,
but no robust confirmation of background lensed sources as they mostly fall outside the 2\arcsec\ SDSS fibers.

Our independent selection has moderate overlap with galaxy-scale lenses in SuGOHI \citep[similar to the comparison in KiDS/GAMA
  from][]{knabel20}. On the one hand, by relying on YattaLens, an algorithm combining lens light subtraction, arc-finding, and lens
modeling \citepalias{sonnenfeld18}, SuGOHI could be more efficient at finding lenses with blended components than our analysis of
brute $gri$ cutouts. On the other hand, our approach classifies a large catalog from PDR2, while SuGOHI have either focused
on spectroscopically confirmed LRGs \citepalias{sonnenfeld18,wong18}, or have used S17A release that covers a 35\% smaller area with
full-color full-depth imaging than PDR2 \citepalias{sonnenfeld20}. Our newly discovered candidates exhibit both extended arcs and simple
double or quad configurations and are not drastically different from those in SuGOHI. Lensed sources mostly have blue colors, and our
visual inspection tends to preferentially retain brighter sources. While the vast majority of lenses are isolated LRGs, small compact
groups also contribute in a few cases. Quantitative properties from lens modeling will be presented in a forthcoming paper.

This analysis paves the way for limiting human inspection in future lens searches not only with LSST, but also with Euclid and Roman.
Unsupervised machine learning has not yet reached the performance of supervised CNNs for lens search, but the results are promising,
especially for identifying peculiar lens configurations that could be omitted in human-assisted training sets \citep{cheng19}. In the
future, combining the two approaches could therefore help increase completeness and purity.

\section*{Acknowledgements}

We would like to thank the referee for useful comments that helped improve the paper. We thank D. Sluse for useful feedback about this
work. RC, SS and SHS thank the Max Planck Society for support through the Max Planck Research Group for SHS. This project has received
funding from the European Research Council (ERC) under the European Union's Horizon 2020 research and innovation program (LENSNOVA:
grant agreement No 771776). YS acknowledges support from the Max Planck Society and the Alexander von Humboldt Foundation in the
framework of the Max Planck$-$ Humboldt Research Award endowed by the Federal Ministry of Education and Research. Also supported by
Riset ITB 2021 (ATJ).

This paper is based on data collected at the Subaru Telescope and retrieved from the HSC data archive system, which is operated by
Subaru Telescope and Astronomy Data Center at National Astronomical Observatory of Japan. The Hyper Suprime-Cam (HSC) collaboration
includes the astronomical communities of Japan and Taiwan, and Princeton University. The HSC instrumentation and software were
developed by the National Astronomical Observatory of Japan (NAOJ), the Kavli Institute for the Physics and Mathematics of the
Universe (Kavli IPMU), the University of Tokyo, the High Energy Accelerator Research Organization (KEK), the Academia Sinica
Institute for Astronomy and Astrophysics in Taiwan (ASIAA), and Princeton University. Funding was contributed by the FIRST
program from Japanese Cabinet Office, the Ministry of Education, Culture, Sports, Science and Technology (MEXT), the Japan
Society for the Promotion of Science (JSPS), Japan Science and Technology Agency (JST), the Toray Science Foundation, NAOJ,
Kavli IPMU, KEK, ASIAA, and Princeton University. This paper makes use of software developed for the LSST. We thank the LSST
Project for making their code available as free software at http://dm.lsst.org.

\bibliography{lenssearch}

\begin{appendix}

\section{Complete list of candidates}

\begin{figure*}
\centering
\includegraphics[width=.9\textwidth]{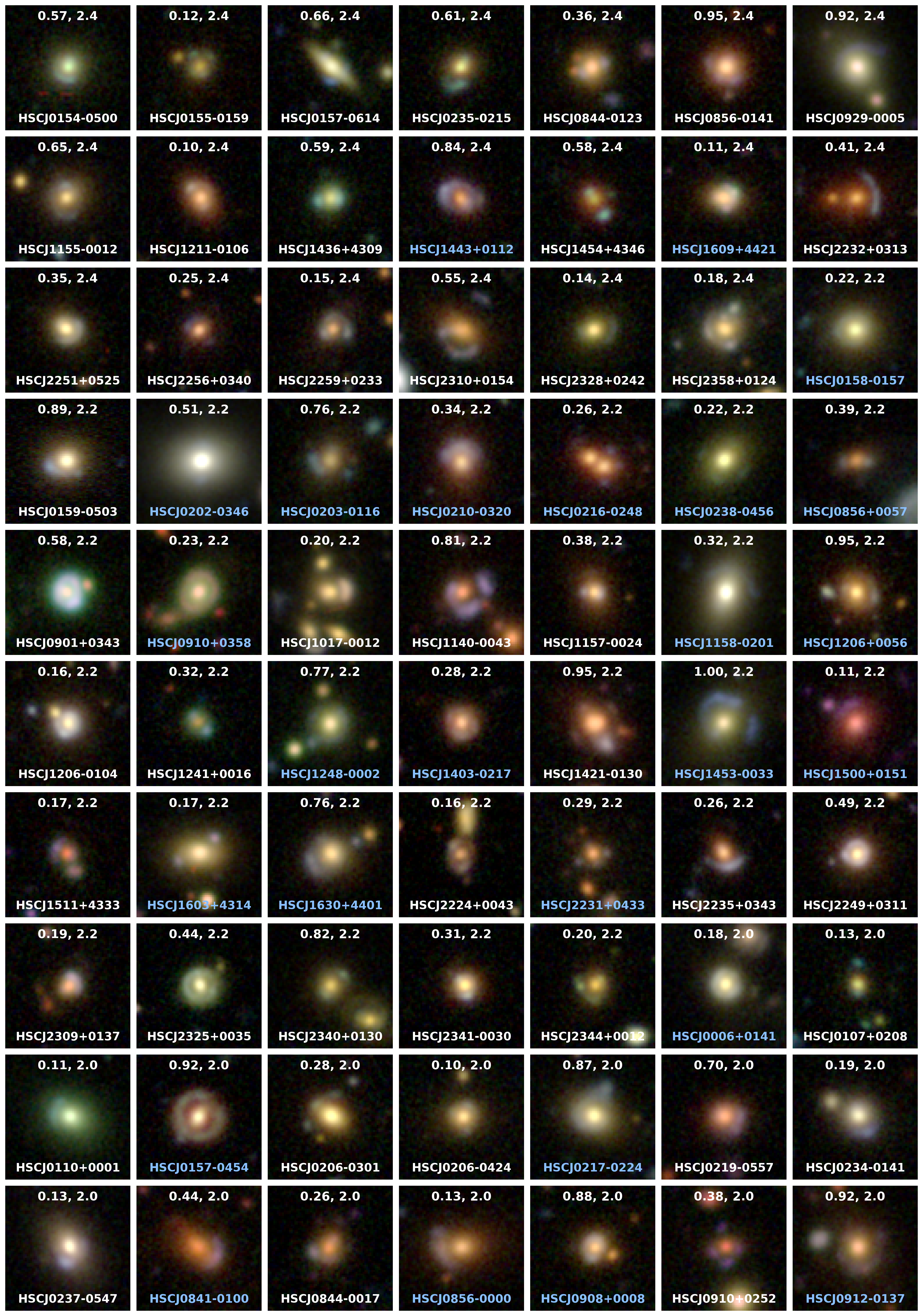}
\caption{HSC three-color $gri$ postage stamps of our grade B lens candidates with $1.5 \leq G < 2.5$. The same format as in
  Fig.~\ref{fig:cnncand} is used.}
\label{fig:cnncandb1}
\end{figure*}

\begin{figure*}
\centering
\includegraphics[width=.9\textwidth]{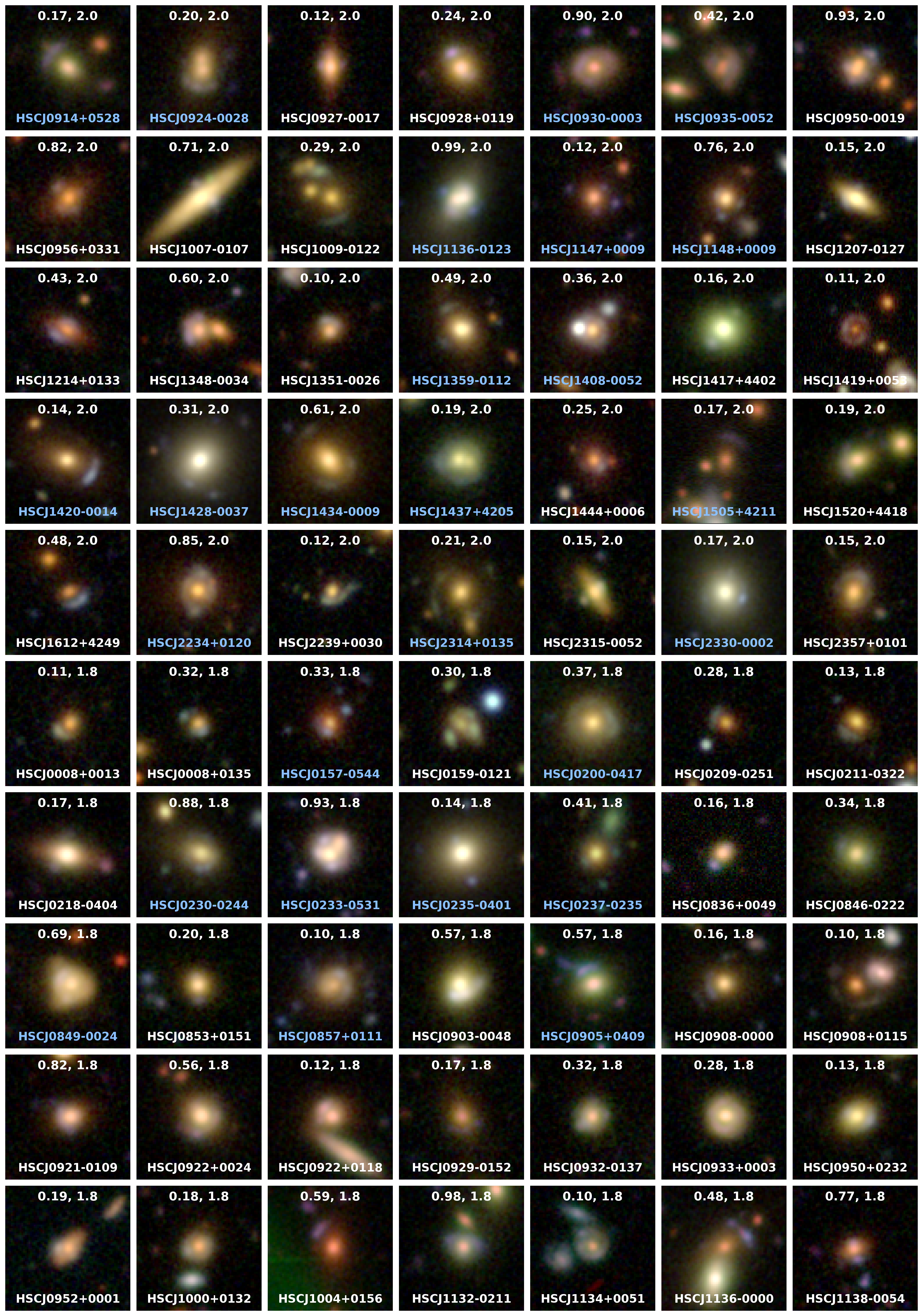}
\caption{continued.}
\label{fig:cnncandb2}
\end{figure*}

\begin{figure*}
\centering
\includegraphics[width=.9\textwidth]{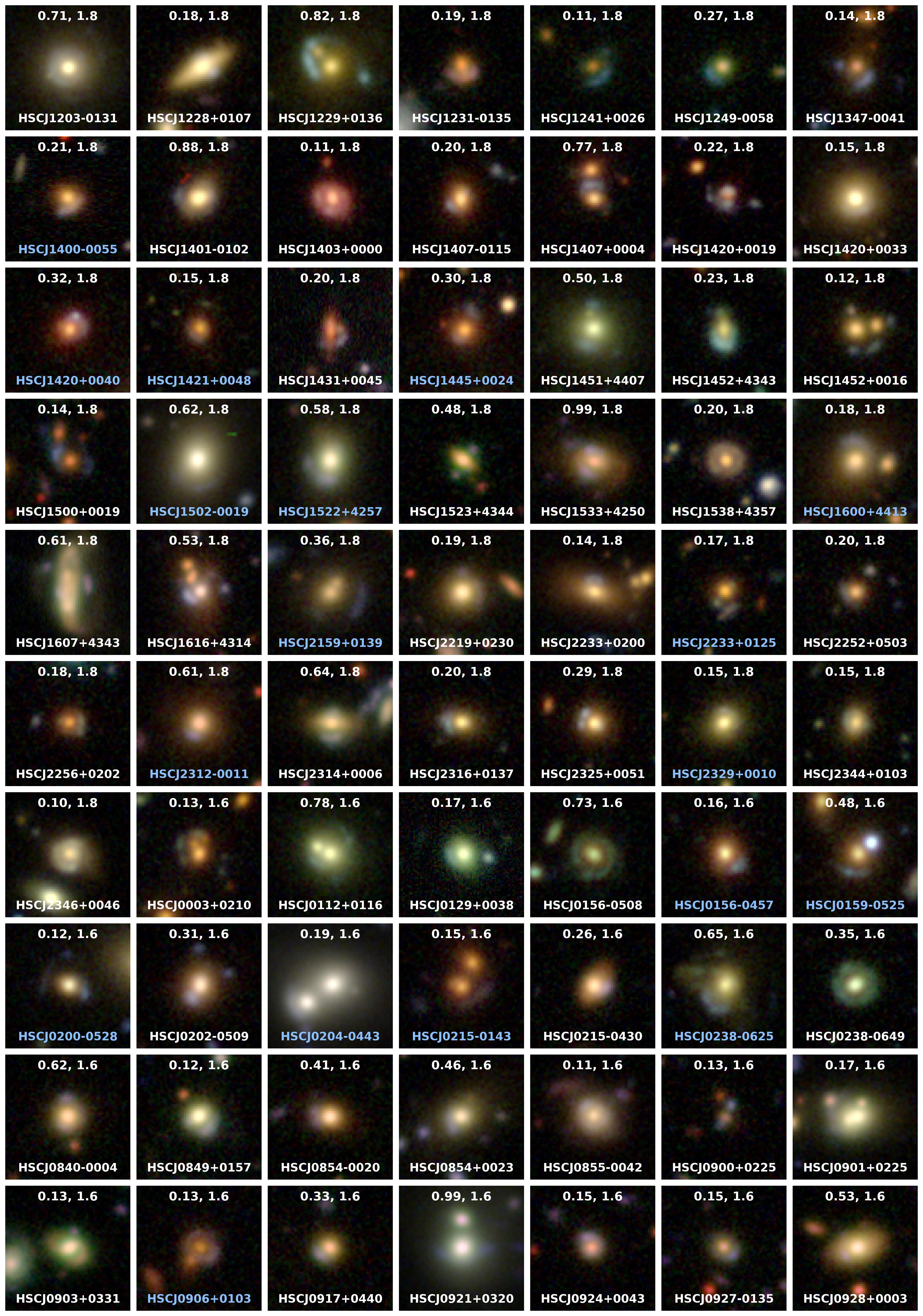}
\caption{continued.}
\label{fig:cnncandb3}
\end{figure*}

\begin{figure*}
\centering
\includegraphics[width=.9\textwidth]{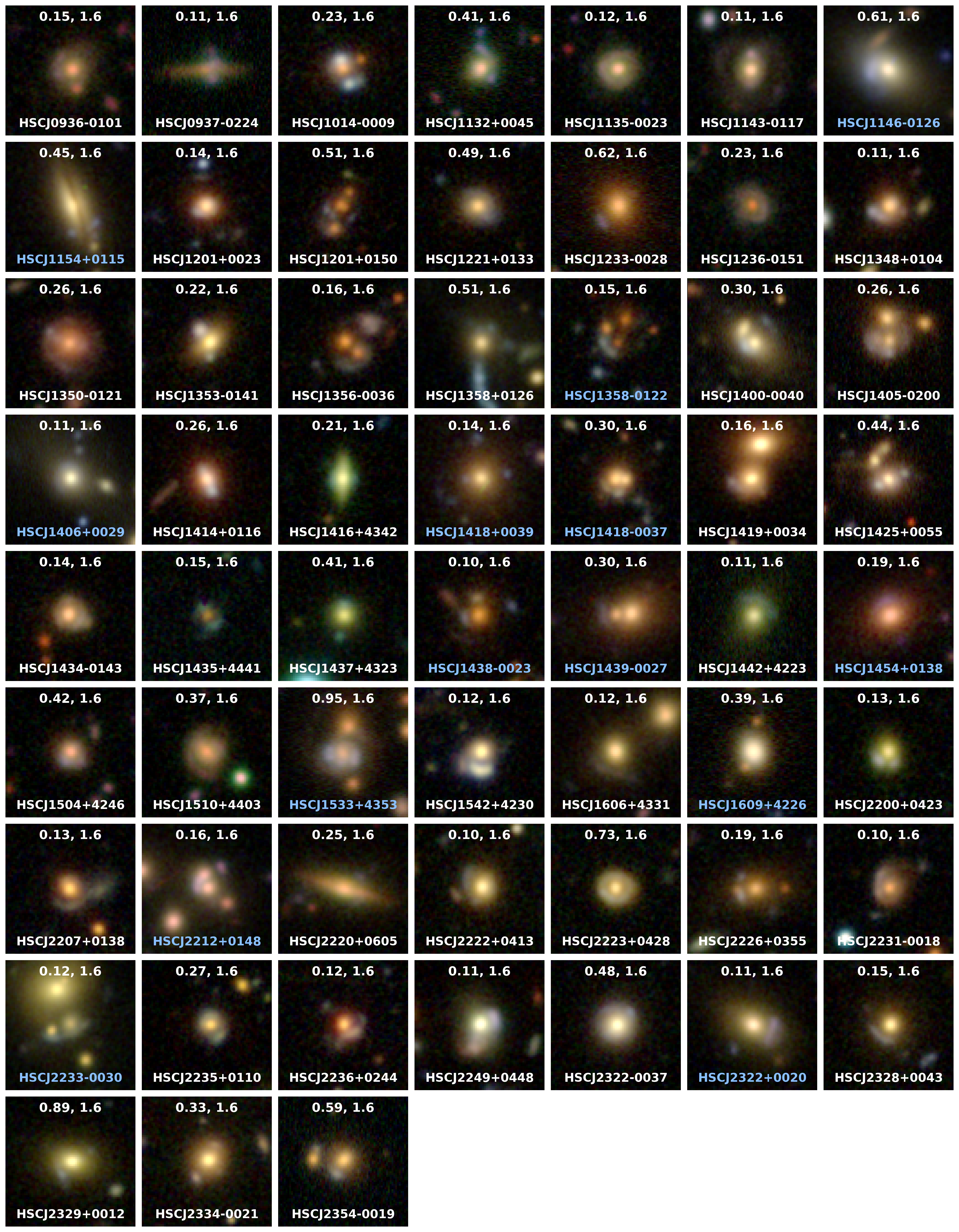}
\caption{continued.}
\label{fig:cnncandb4}
\end{figure*}

\end{appendix}

\end{document}